\documentstyle[aps,prbbib,twocolumn,epsf]{revtex}

\begin{document}
\draft
\title{ Statistical correlation for the composite Boson}

\author{Baigeng Wang and Jian Wang$^{a)}$}
\address{Department of Physics, The University of Hong Kong, 
Pokfulam Road, Hong Kong, China\\
}
\maketitle

\begin{abstract}
It is well known that the particles in a beam of Boson obeying 
Bose-Einstein statistics tend to cluster (bunching effect), while 
the particles in a degenerate beam of Fermion obeying 
Fermi-Dirac statistics expel each other (anti-bunching
effect). Here we investigate, for the first time, the statistical
correlation effect for the composite Boson, which is formed from a spin
singlet entangled electron pair. By using nonequilibrium Green's function
technique, we obtain a positive cross correlation for this kind of the
composite Boson when the external voltage is smaller than the gap energy, 
which demonstrates that a spin singlet entangled electron
pair looks like a composite Boson. In the larger voltage limit, 
the cross correlation becomes negative due to the contribution of the
quasiparticles. At large voltages, the oscillation between Fermionic
and Bosonic behavior of cross correlation is also observed in the strong
coupling regime as one changes the position of the resonant levels. 
Our result can be easily tested in a three-terminal 
normal-superconductor-superconductor (N-S-S) hybrid mesoscopic system.

\end{abstract}

\pacs{74.50.+r, 72.70.+m, 74.40.+k, 73.23.-b}

\section{Introduction}

There are two kinds of quantum statistics in nature. All particles have
either half-integral or integral spin (in units of the Plank constant 
$\hbar $) and they obey Fermi-Dirac\cite{fermi} or 
Boson-Einstein\cite{bose} statistics respectively. It is also 
noted\cite{landau} that there is an effective attraction between
the Bosons and an effective repulsion between the Fermions. This is
the well known statistical correlation effects\cite{buttiker1}, which are 
purely quantum effect. The experiments examining the quantum statistical 
properties date back to the pioneer work, by Hanbury Brown and 
Twiss (HBT)\cite{brown}. They used photon intensity interferometry 
to probe the intensity correlation information between two partial 
beams, which was generated by a beam splitter. Due to the Bosonic 
property of photon, the positive intensity correlation was observed, 
indicating an enhanced probability for the simultaneous detection of 
two photons, one in each partial beam. This means
that photons tend to bunch in cluster. Several theoretical works have
suggested the different analogies of this experiment with electrons in
mesoscopic systems. The Fermionic analog of HBT experiments, one by 
Henny et al.\cite{henny} and the other by Oliver et al.\cite{oliver}, 
showed the expected negative intensity correlation and observed the 
anti-bunching effect. In this paper we will investigate the HBT experiment 
for the composite Boson. This composite Boson is formed by a spin 
singlet entangled electron pair, which will be discussed below. Due 
to the zero total spin for the entangled electron pair, we expect that
these composite Bosons tend to bunch in cluster. Motivated by the
application in quantum communication and computation, Burkard et 
al\cite{burkard} have studied entangled electrons in an interacting 
many-body environment. They found that the Fano factor for singlets is 
twice as large as for independent classical particles and is reduced to 
zero for triplets. Torri\`{e}s and Martin\cite{torries} investigated a 
three-terminal N-N-S mesoscopic system, both positive and negative 
correlations were found in the Andreev regime. Very recently, 
Samueisson and B\"{u}ttiker\cite{samu} studied the same structure and 
found the positive correlation for a wide range of junction parameters 
which survives even in the absence of the proximity effect. The
statistics of charge transport of a three terminal N-N-S beam splitter
has also been investigated\cite{borlin} and positive cross correlation
is found between the currents in two normal leads for a wide parameter
range. Instead of the structures of the 
references\onlinecite{torries,samu,borlin}, here we consider 
a three-terminal mesoscopic N-S-S hybrid system. This structure is a 
direct photon analogy of the HBT interferometer which has a normal 
lead and two superconducting leads. A quantum dot, connected by these 
three terminals, acts as a splitter. Suppose that the chemical 
potentials 
$\mu_s$ for both superconducting reservoirs are set to zero, and the 
chemical potential for the normal is above zero, i.e., $eV>0$, which 
guarantees the electron current passing from the normal lead to 
both superconducting leads. We further assume the temperature is very low.
If the external voltage $eV$ is smaller than the gap energy $\Delta$ of
the superconducting leads, the single quasiparticle current is forbidden.
In this case, we only have two-electron current due to the presence of 
Andreev reflection process, i.e., incoming electrons being Andreev 
reflected into outgoing holes with the transfer of a Cooper pair into the 
superconductor. This means that an electron (with
energy $\epsilon$ above the Fermi level and spin $\sigma$) in the
normal lead, has to combine another electron (with energy $-\epsilon$, 
below the Fermi level and spin $-\sigma$) to pass through the NS interface.
Does this entangled electron pair\cite{lesovik} look like a composite Boson? 
or rather, can we obtain a positive cross correlation function 
($<\!\Delta I_\alpha \Delta I_\beta \!>$ with $\alpha \neq \beta$) between
two superconducting leads? The purpose of this paper is to answer this 
question. We note that due to the current conservation the cross 
correlation function of a two-lead system must be negative regardless of
normal or superconducting leads. Instead of considering the fluctuation 
in a single electron beam through the two-lead system, the HBT experiment 
considered here focuses on the cross correlation of two beams from
the beam splitter. Hence we expect positive cross correlation at small
voltages which is indeed what we found in this work. When $eV>\Delta$, 
the quasiparticles will also participate the transport. Due to the 
Fermionic nature of quasiparticles, it will partially cancel the positive 
contribution of the entangled electron pair to the cross correlation. 
The competition of these two contributions from the entangled electron 
pair and quasiparticles can lead to either positive or negative cross 
correlation depending on which contribution dominates.

\section{Theoretical formulation}

We begin with the following model Hamiltonian
\begin{eqnarray}
&&H =\sum_{p}\epsilon_{p}C_{1,p\sigma }^{+}C_{1,p\sigma
}+\sum_{kn}[\sum_\sigma \epsilon_{k}C_{n,k\sigma }^{+}C_{n,k\sigma } 
\nonumber \\
&&+\Delta C_{n,k\uparrow}^{+}C_{n,-k\downarrow }^{+} 
+\Delta C_{n,-k\downarrow }C_{n,k\uparrow }]
+\sum_{\sigma }\epsilon_{0} d_{\sigma }^{+}d_{\sigma } 
\nonumber \\
&&+\sum_{p\sigma }[T_{1,p}C_{1,p\sigma }^{+}d_{\sigma }+c.c.] 
+\sum_{kn\sigma}[T_{n,k}C_{n,k\sigma }^{+}d_{\sigma }+c.c.]
\label{ham}
\end{eqnarray}
where the first term denotes the Hamiltonian of the normal lead. The second 
term ($n=2,3$) describes the Hamiltonian of two BCS superconducting leads. 
Here $C_{1,k\sigma}^{\dagger}$ is the creation operator 
of electrons in the normal lead and $C_{n,k\sigma}^{\dagger}$ is the 
corresponding creation operator in the superconducting lead. The 
third term is the Hamiltonian for a quantum dot, which is used to 
mimic a tunable beam splitter. Here we have applied a gate voltage which
can control the level of the dot so that $\epsilon_0 = \epsilon_0^{(0)}
+ev_g$. Without loss of generality, we set $\epsilon_0^{(0)}=0$.
The other terms in Eq.(\ref{ham}) are Hamiltonians describing 
the couplings between the quantum dot and leads.
To simplify the discussion, we have assumed that two superconducting 
leads have the same gap energy $\Delta$. We have also neglected the 
supercurrent between two superconducting leads\cite{nagaev} and assumed 
that the hopping matrix elements are independent of the spin index.

In the following, we will calculate the cross correlation between two 
partial beams through two superconducting leads.
The current operator for the superconducting lead 2 or 3 is 

$$\hat{I}_\alpha=\hat{I}_{\alpha\uparrow}(t)
+\hat{I}_{\alpha\downarrow}(t)$$
with 
\[
\hat{I}_{\alpha\sigma}(t)=ie[\sum_k C^{\dagger}_{\alpha,k\sigma}
C_{\alpha,k\sigma}, H] = ie\sum_{k}[T_{\alpha k}C_{\alpha,k\sigma
}^{+}d_{\sigma }-c.c.]
\]
where $\alpha=2,3$. Due to the electron-hole symmetry of the system, 
we have $\hat{I}_{\alpha\uparrow}(t)=\hat{I}_{\alpha\downarrow}(t)$.
Hence the current operator can be rewritten as
\[
\hat{I}_{\alpha}(t)=2ie\sum_{k}[T_{\alpha k}C_{\alpha,k\uparrow}^{+}
d_{\uparrow}-c.c.]
\]
The cross correlation between two superconducting leads
is defined as
\[
P_{23}\equiv<\!\Delta I_{2}(t_{1})\Delta I_{3}(t_{2})\!>
\equiv <\![\hat{I}_{2}(t_{1})-\bar{I}
_{2}][\hat{I}_{3}(t_{2})-\bar{I}_{3}]\!>
\]
with $\bar{I}_\alpha \equiv <{\hat I}_\alpha>$.
Here $<...>$ denotes both the statistical average and quantum average
on the nonequilibrium state. Using the expression of the current 
operator, the cross correlation between two superconducting leads is

\begin{eqnarray}
&&P_{23}=-4e^{2}\sum_{k,k^{\prime }} 
[T_{2,k}T_{3,k^{\prime }} G_{d\uparrow k\uparrow}^{<}(2,1) 
G_{d\uparrow k^{\prime }\uparrow }^{>}(1,2) 
\nonumber \\
&&+T_{2,k}^{\ast }T_{3,k^{\prime }}^{\ast}G_{k^{\prime }\uparrow 
d\uparrow }^{<}(2,1)G_{k\uparrow d\uparrow}^{>}(1,2) \nonumber \\
&&-T_{2,k}T_{3,k^{\prime }}^{\ast } 
G_{k^{\prime }\uparrow k\uparrow
}^{<}(2,1)G_{d\uparrow d\uparrow }^{>}(1,2) \nonumber \\
&&-T_{2,k}^{\ast}T_{3,k^{\prime }}G_{d\uparrow d\uparrow }^{<}(2,1)
G_{k\uparrow k^{\prime }\uparrow }^{>}(1,2)]
\end{eqnarray}
where we have used abbreviation $G(t_1,t_2) = G(1,2)$ and we have used
$k$ and $k'$ to label, respectively, the second and third superconducting 
lead. The Green's functions ${\bf G}^{r,a,<,>}$ in 2 $\times $ 2 Nambu 
representation take the following forms\cite{cuevas,sun1,sun2} 

\begin{eqnarray}
&&G_{\alpha \beta }^{r,a}(t_{1},t_{2})=\mp i\theta (\pm t_{1}\mp t_{2})
\nonumber \\
&\times& \left(
\begin{tabular}{ll}
$<\!\{X_{\alpha \uparrow }(t_{1}),Y_{\beta \uparrow }^{+}(t_{2})\}\!>$ & 
$<\!\{X_{\alpha \uparrow }(t_{1}),Y_{\beta \downarrow }(t_{2})\}\!>$ \\ 
$<\!\{X_{\alpha \downarrow}^{+}(t_{1}),Y_{\beta \uparrow}^{+}(t_{2})\}\!>$ 
& $<\!\{X_{\alpha \downarrow}^{+}(t_{1}),Y_{\beta \downarrow}(t_{2})\}\!>$
\end{tabular}
\right) \nonumber 
\end{eqnarray}

\[
G_{\alpha \beta }^{<}(t_{1},t_{2})=i\left(
\begin{tabular}{ll}
$<\!Y_{\beta \uparrow }^{+}(t_{2})X_{\alpha \uparrow }(t_{1})\!>$ & 
$<\!Y_{\beta \downarrow }(t_{2})X_{\alpha \uparrow }(t_{1})\!>$ \\ 
$<\!Y_{\beta \uparrow }^{+}(t_{2})X_{\alpha \downarrow }^{+}(t_{1})\!>$ & 
$<\!Y_{\beta \downarrow }(t_{2})X_{\alpha \downarrow }^{+}(t_{1})\!>$
\end{tabular}
\right) 
\]

\begin{center}
\[
G_{\alpha \beta }^{>}(t_{1},t_{2})=-i\left(
\begin{tabular}{ll}
$<\!\!X_{\alpha \uparrow }(t_{1})Y_{\beta \uparrow }^{+}(t_{2})\!\!>$ & 
$<\!\!X_{\alpha \uparrow }(t_{1})Y_{\beta \downarrow }(t_{2})\!\!>$ \\ 
$<\!\!X_{\alpha \downarrow}^{+}(t_1)Y_{\beta \uparrow}^{+}(t_2)\!\!>$ & 
$<\!\!X_{\alpha \downarrow}^{+}(t_{1})Y_{\beta \downarrow }(t_2)\!\!>$
\end{tabular}
\right) 
\]
\end{center}
where X and Y stand for the annihilation operators, such as 
$C_{1,p}$, $C_{n,k}$, and $d$. These Green's functions satisfy 
the general relation, ${\bf G}^{>}={\bf G}^{<}+{\bf G}^{r}-{\bf G}^{a}$. 
Using the Keldysh equation\cite{rammer}

\[
{\bf G}^{<,>}=(1+{\bf G}^{r} {\bf \Sigma}^{r}) {\bf G}_{0}^{<,>}
(1+{\bf \Sigma}^{a} {\bf G}^{a})+{\bf G}^{r}{\bf \Sigma}^{<} {\bf G}^{a} 
\]

we have the following relations
\begin{eqnarray}
&&G_{d\uparrow k\uparrow }^{<,>}(t_{1},t_{2}) =T_{2,k}^{\ast }\int
dt~[G_{d\uparrow d\uparrow }^{r}(t_{1},t)g_{k\uparrow k\uparrow
}^{<,>}(t,t_{2}) \nonumber \\
&&+G_{d\uparrow d\uparrow }^{<,>}(t_{1},t)g_{k\uparrow
k\uparrow }^{a}(t,t_{2}) 
+G_{d\uparrow d\downarrow }^{r}(t_{1},t)g_{k\downarrow k\uparrow
}^{<,>}(t,t_{2}) \nonumber \\
&&+G_{d\uparrow d\downarrow }^{<,>}(t_{1},t)g_{k\downarrow
k\uparrow }^{a}(t,t_{2})]
\end{eqnarray}

\begin{eqnarray}
&&G_{k\uparrow d\uparrow }^{<,>}(t_{1},t_{2}) =T_{2,k}\int dt~
[g_{k\uparrow
k\uparrow }^{<,>}(t_{1},t)G_{d\uparrow d\uparrow }^{a}(t,t_{2})
\nonumber \\
&&+g_{k\uparrow
k\uparrow }^{r}(t_{1},t)G_{d\uparrow d\uparrow }^{<,>}(t,t_{2}) 
+g_{k\uparrow k\downarrow }^{<,>}(t_{1},t)G_{d\downarrow d\uparrow
}^{a}(t,t_{2})
\nonumber \\
&&+g_{k\uparrow k\downarrow }^{r}(t_{1},t)G_{d\downarrow
d\uparrow }^{<,>}(t,t_{2})]
\end{eqnarray}

\begin{eqnarray}
&&G_{k\uparrow k^{\prime }\uparrow }^{<,>}(t_{1},t_{2}) =T_{3,k^{\prime
}}^{\ast }\int dt~[G_{k\uparrow d\uparrow }^{r}(t_{1},t)g_{k^{\prime
}\uparrow k^{\prime }\uparrow }^{<,>}(t,t_{2})
\nonumber \\
&&+G_{k\uparrow d\uparrow
}^{<,>}(t_{1},t)g_{k^{\prime }\uparrow k^{\prime }\uparrow }^{a}(t,t_{2}) 
+G_{k\uparrow d\downarrow }^{r}(t_{1},t)g_{k^{\prime }\downarrow k^{\prime
}\uparrow }^{<,>}(t,t_{2})
\nonumber \\
&&+G_{k\uparrow d\downarrow }^{<,>}(t_{1},t)g_{k^{\prime }\downarrow 
k^{\prime }\uparrow }^{a}(t,t_{2})]
\end{eqnarray}
where $G^r_{k\uparrow,d\sigma}$ is given by
\begin{eqnarray}
&&G_{k\uparrow d\sigma}^{r}(t_{1},t_{2}) =T_{2,k}\int dt[g_{k\uparrow
k\uparrow }^{r}(t_{1},t)G_{d\uparrow d\sigma}^{r}(t,t_{2})
\nonumber \\
&&+g_{k\uparrow k\downarrow }^{r}(t_{1},t)G_{d\downarrow d\sigma
}^{r}(t,t_{2})]
\end{eqnarray}

Substituting the above relations to Eq.(1) and taking the Fourier
transform, we obtain
\begin{eqnarray}
&&P_{23}=-4e^{2}\Gamma _{2}\Gamma _{3}\int 
\frac{dE }{2\pi } \{({\bf G}^{r} {\bf g}^{<}
+{\bf G}^{<} {\bf g}^{a})_{\uparrow \uparrow}
({\bf G}^{r} {\bf g}^{>}
\nonumber \\
&&+{\bf G}^{>} {\bf g}^{a})_{\uparrow \uparrow}
+({\bf g}^{<} {\bf G}^{a} +{\bf g}^{r} {\bf G}^{<})_{\uparrow \uparrow}
({\bf g}^{>} {\bf G}^{a}
+{\bf g}^{r} {\bf G}^{>})_{\uparrow \uparrow}
\nonumber \\
&&-G^>_{\uparrow \uparrow}
\{({\bf g}^{r} {\bf G}^{r} {\bf g}^{<})_{\uparrow \uparrow}
+[({\bf g}^{<} {\bf G}^{a}+{\bf g}^{r} {\bf G}^{<}) 
{\bf g}^{a}]_{\uparrow \uparrow} \} \nonumber \\
&&-G_{\uparrow \uparrow}^{<}\{({\bf g}^{r} {\bf G}^{r} {\bf g}^{>} 
)_{\uparrow \uparrow} +[({\bf g}^{>} {\bf G}^{a}+{\bf g}^{r} {\bf G}^{>})
{\bf g}^{a}]_{\uparrow \uparrow}
\}\}
\label{final}
\end{eqnarray}
where $\Gamma_{\alpha}=2\pi \sum_k \rho_{N\alpha}|T_{\alpha k}|^2$ with
$\alpha=2,3$ are the linewidth functions. Here $\rho_{N2,3}$ are the normal 
density of states of the superconducting leads 2 and 3. We have used the 
wide-band limit\cite{jauho} and thus the linewidth function is independent 
of the energy. $G_{\sigma \sigma^{\prime }}^{r,a,<,>} \equiv G_{d\sigma 
d\sigma^{\prime}}^{r,a,<,>}$ are the full Green's functions for the
quantum dot in presence of the leads, while $g_{\sigma \sigma^{\prime
}}^{r,a,<,>}$ are the exact Green's functions for BCS superconductor in the
absence of the coupling between the leads and quantum dot. Eq.(\ref{final})
is the central result of this paper. It describes the cross 
correlation for a three-terminal hybrid N-S-S system and is valid at any 
temperature and finite voltage, i.e., valid for both $eV \geq \Delta$
and $eV < \Delta$. In order to calculate this 
correlation, one must know all the Green's functions. The exact 
Green's functions $g_{\sigma \sigma ^{\prime }}^{r,a,<}$ for the isolated 
superconducting leads are\cite{wang1,foot1}

\[
{\bf g}^{r}(E)=-\frac{i\zeta (E)}{2\sqrt{E^{2}-\Delta^{2}}}
\left(
\begin{tabular}{ll}
$E$ & $\Delta $ \\ 
$\Delta $ & $E$
\end{tabular}
\right)=[{\bf g}^{a}(E)]^{+} 
\]

\begin{center}
\[
{\bf g}^{<}(E)=if(E)\theta (|E| -\Delta)\frac{\zeta (E)}{\sqrt{E^{2}-\Delta^2}}
\left(
\begin{tabular}{ll}
$E$ & $\Delta $ \\ 
$\Delta $ & $E$
\end{tabular}
\right) 
\]
\end{center}
where $f(E)=1/[\exp(\beta (E-E_F)+1]$ is the well known Fermi distribution
function, $\theta (x)$ is the step function and $\zeta (E)=1$ when $
E>-\Delta $, otherwise $\zeta (E)=-1$. We will choose the Fermi energy of the
normal lead in line with the chemical potential $\mu_s$ of superconducting 
condensate which is set to zero, i.e., $E_F=\mu_s=0$. The retarded Green's
function for the quantum dot can be calculated using the Dyson equation
\[
{\bf G}^{r}(E)=\frac{1}{[{\bf G}_{0}^{r}(E)]^{-1}-{\bf \Sigma}^{r}(E)} 
\]
with
\[
{\bf G}_{0}^{r}(E)=\frac{1}{\left(
\begin{tabular}{cc}
$E-\epsilon _{0}$ & $0$ \\ 
$0$ & $E+\epsilon _{0}$
\end{tabular}
\right)} 
\]
and
\[
{\bf \Sigma}^{r}(E)=-\frac{i}{2}\Gamma _{1}\left(
\begin{tabular}{ll}
$1$ & $0$ \\ 
$0$ & $1$
\end{tabular}
\right)-\frac{i}{2}(\Gamma _{2}+\Gamma _{3})\frac{\zeta (E)}
{\sqrt{E^{2}-\Delta^{2}}}\left(
\begin{tabular}{ll}
$E$ & $\Delta $ \\ 
$\Delta $ & $E$
\end{tabular}
\right) 
\]
The lesser Green's function can be obtained from the Keldysh equation 
${\bf G}^<={\bf G}^r{\bf \Sigma}^< {\bf G}^a$. Here the lesser 
self-energy is given by
\begin{eqnarray}
&&{\bf \Sigma}^{<}(E) =i\Gamma _{1}\left(
\begin{tabular}{cc}
$f(E+eV)$ & $0$ \\ 
$0$ & $f(E-eV)$
\end{tabular}
\right)+ 
\nonumber \\
&&if(E)\theta (|E|-\Delta)\frac{\Gamma_{2}+\Gamma_{3}}{
{\sqrt{E^{2}-\Delta ^{2}}}}
\zeta (E)
\left(
\begin{tabular}{ll}
$E$ & $\Delta $ \\ 
$\Delta $ & $E$
\end{tabular}
\right)
\nonumber 
\end{eqnarray}
Let's first consider the case that external voltage is smaller than the 
gap energy and consider zero temperature behavior so that there is no 
quasiparticles participating the transport. In this case, only 
two-electron current exists, i.e., the current from
incoming electron and Andreev reflected hole, we have 
${\bf g}^r={\bf g}^a$ and ${\bf g}^{<,>}=0$. 
using the fact that 

\begin{equation}
{\bf G}^< = i \Gamma_1 {\bf G}^r
\left(
\begin{tabular}{cc}
$f_+$ & 0 \\ 
0 & $f_-$
\end{tabular}
\right)
{\bf G}^a
\end{equation}
and 
\begin{equation}
{\bf G}^> = i \Gamma_1 {\bf G}^r
\left(
\begin{tabular}{cc}
$f_+-1$ & 0 \\ 
0 & $f_- -1$
\end{tabular}
\right)
{\bf G}^a
\end{equation}
Eq.(\ref{final}) can be further simplified as
\begin{eqnarray}
&&P_{23}= \frac{e^{2} \Gamma_1^2 \Gamma _{2}\Gamma _{3}\Delta^2}
{\Delta^2-E^2} \int \frac{dE }{2\pi } f_{-} (1-f_+) \nonumber \\
&& \times |G^r_{\uparrow \uparrow} G^a_{\downarrow \downarrow} 
- G^r_{\uparrow \downarrow} G^a_{\uparrow \downarrow}|^2 \nonumber \\
&&=\frac{4e^{2} \Gamma _{2}\Gamma _{3}}{(\Gamma_2+\Gamma_3)^2}
\int \frac{dE }{2\pi } f_{-} (1-f_+) \nonumber \\
&& \times T_A(E)(1-T_A(E))
\label{final1} 
\end{eqnarray}
where $T_A(E) = \Gamma_1^2 G^r_{\uparrow \downarrow} G^a_{\uparrow 
\downarrow}$ is the Andreev reflection coefficient
and $f_{\pm}(E) = f(E\pm eV)$. 
Just as we expected, Eq.(\ref{final1}) is a positive quantity.
To get more physical insight, we will
assume that $eV$ are small enough and we will keep only the first order
in $V$ in Eq.(\ref{final1}). 
We have 
\begin{eqnarray}
&&P_{23}=\frac{\Gamma _{1}^{2}\Gamma _{2}\Gamma
_{3}e^{3}V}{\pi[\epsilon _{0}^{2}+\frac{\Gamma _{1}^{2}}{4}+\frac{(\Gamma
_{2}+\Gamma _{3})^{2}}{4}]^{4}} \nonumber \\
&&\times\{\epsilon _{0}^{4}+\frac{\epsilon
_{0}^{2}[\Gamma _{1}^{2}+(\Gamma _{2}+\Gamma _{3})^{2}]}{2}+\frac{[\Gamma
_{1}^{2}-(\Gamma _{2}+\Gamma _{3})^{2}]^{2}}{4}\} 
\end{eqnarray}
For $eV >\Delta$, we have to calculated $P_{23}$ numerically which is
presented in the next section. 

\section{Result and discussion}

We first use Eq.(\ref{final1}) to calculate the cross correlation at finite
voltage while keeping $eV<\Delta$. In the following, we will use $\Delta$ 
as the unit of energy and
study the symmetric case where $\Gamma_2=\Gamma_3$. 
In Fig.1 we show the cross correlation versus the gate voltage at fixed
external bias $eV=0.6$. Four different sets of coupling constants
$\Gamma$ are chosen: (1). $\Gamma_1=0.8$ and $\Gamma_2=0.8$. In this 
case, it represents the strong coupling between
leads and the quantum dot. The cross correlation (solid line)
displays two broad peaks located symmetrically at $ev_g=\pm 0.6$. 
(2). $\Gamma_1=0.8$ and $\Gamma_2=0.1$. In this case, the normal
lead couples strongly with the quantum dot while the superconducting
leads couple weakly. The cross correlation (dotted line) exhibits a single 
peak at $v_g=0$. (3). $\Gamma_1=0.1$ and $\Gamma_2=0.8$. This 
is the reverse of case (2). We see that the cross correlation (dot-dashed 
line) shows a flat region near $v_g=0$. (4). $\Gamma_1=0.1$ and 
$\Gamma_2=0.1$. This is the weak coupling case. The cross 
correlation (dashed line) has two sharp peaks close to $v_g=0$ and 
decays quickly away from it. To understand these features, we notice that 
two terms ($F_1=\int dE ~ T_A(E)$ and $F_2=\int dE ~ T_A^2(E)$) in 
Eq.(\ref{final1}) tend to cancel each other. In the strong coupling case, 
the contribution from both terms are of the same order of magnitude. 
Both show broad peak near $v_g=0$ with the second term decreasing faster 
away from $v_g=0$ (see Fig.2a). As a result, we obtain the double-peak 
structure as shown in Fig.1. When $\Gamma_1=0.8$ and $\Gamma_2=0.1$, 
the contribution from $F_2$ is much smaller than that 
of $F_1$ and hence just one peak shows up (Fig.2b). For $\Gamma_1=0.1$
and $\Gamma_2=0.8$, the addition of $F_1$ and $F_2$ gives a long plateau
between $ev_g=-0.5$ and $ev_g=0.5$ (Fig.2c).  In the weak coupling regime, 
$F_1$ and $F_2$ give comparable contributions with a single peak. Similar 
to the strong coupling case, the integral of $F_2$ decreases faster than 
that of $F_1$ resulting again a double-peak structure (Fig.2d). As one 
decreases the external bias to $eV=0.2$, the cross correlation in the 
strong coupling case still shows double peak structure with smaller 
amplitude (Fig.3). We also find that the peak position is shifted 
towards origin and peak to valley ratio becomes much smaller. In the 
weak coupling regime, the cross correlation is roughly unchanged. In the 
case of $\Gamma_1=0.8$ and $\Gamma_2=0.1$ (or vice versa), the cross 
correlation decreases. Now we examine the cross correlation versus 
external bias at fixed energy levels. Fig.4a displays the cross correlation
$P_{23}$ versus external voltage when $\epsilon_0=0$. We see that, 
except for 
$\Gamma_1=0.8$ and $\Gamma_2=0.1$ that $P_{23}$ increases monotonically,
$P_{23}$ develops a plateau region for the other three sets of coupling 
parameters. These plateau regions are due to the resonant tunneling which
can be seen from Fig.4b where the differential cross correlation 
$dP_{23}/dV$ versus external voltage is depicted. 
Here we see typical behavior of the shot noise\cite{wei1}: 
a minimum separated by two peaks. The minimum is due to the resonant
Andreev reflection since $dP_{23}/dV \sim T_A (1-T_A)$. As one increases
the energy level ($\epsilon_0=0.3$), the dip between two peaks can no longer
reach zero indicating that the maximum Andreev reflection coefficient $T_A$
is much less than one. We also note that for $\Gamma_1=0.1$ and 
$\Gamma_2=0.8$, only one peak is left and resonant feature disappeared.

To study the effect of quasiparticle when $eV > \Delta$, we calculate 
the cross correlation using Eq.(\ref{final}). Fig.6 shows the cross 
correlation versus external voltage at $\epsilon_0=0.0$. We see that 
once the voltage is larger than the gap energy $\Delta$, the cross 
correlation decreases quickly indicating Fermionic contributions. 
For the strong coupling case, $P_{23}$ becomes negative in the large $V$
limit. This can be understood as follows. When $eV>\Delta$, electrons
with energy less than $eV$ will all participate in transport. In
particular, for incoming electrons with energy inside the superconducting 
gap, only two-electron current is allowed and hence the contribution 
to the cross correlation should be positive as we just discussed above. 
However, when the energy of incoming electron is outside of the gap 
the current comes from of four processes\cite{btk,cuevas}: (1). Andreev 
reflection; (2). the conventional electron tunneling through the system; 
(3). "Branch crossing" process\cite{btk}: an electron incident from the 
normal lead converting into a hole like in the superconducting leads; 
(4). An electron (or a hole) incident from the normal lead tunnels into the 
superconducting lead, picks up a quasiparticle (or a quasihole) in the 
superconducting lead and creates (or annihilates) a Cooper pair. In these
processes, the latter three give negative contributions to the cross
correlation. Competition between Andreev reflection process and the 
rest of three processes give rise either positive or negative cross 
correlation depending on which process dominates (see Fig.6). Typically, 
near the resonance the Breit-Wigner form for the Andreev reflection 
coefficient reads\cite{beenakker,wei2}

\begin{equation}
T_A = \frac{\Gamma_1^2 \Gamma_2^2}{4(E^2 - \epsilon_0^2 + \Gamma\delta 
\Gamma/4)^2 + \Gamma_1^2 \Gamma_2^2 + \epsilon_0^2(\Gamma + 
\delta \Gamma)^2}
\end{equation}
and transmission coefficient for normal tunneling process
\begin{equation}
T = \frac{\Gamma_1 \Gamma_2}{(E-\epsilon_0)^2+\Gamma^2/4}
\end{equation}
where $\Gamma=\Gamma_1+\Gamma_2$ and $\delta \Gamma = \Gamma_1 -
\Gamma_2$. 
We see that the Andreev reflection is suppressed when off resonance. 
Furthermore, at large external bias if the resonant energy is outside 
of the gap, the Andreev reflection is drastically suppressed and
normal tunneling is allowed at certain energy. Therefore, we expect
negative cross correlation in this case. In Fig.7, we depict $P_{23}$
versus $V$ at $\epsilon_0=2.0$. Since the resonant level is outside
of the gap, the plateau region for $P_{23}$ when $eV$ is inside the 
gap disappeared. We see that
except for the case of $\Gamma_1=0.8$ and $\Gamma_2=0.1$, $P_{23}$ 
becomes negative at large voltages. Our numerical result shows that
at even larger $\epsilon_0$, the transport of quasiparticle dominates 
and all $P_{23}$ are negative at large external voltage. Finally, 
we plot in Fig.8 the $P_{23}$ versus $v_g$ at $eV=4$. We see that at 
large voltages, all
the cross correlation functions become negative. For the strong coupling
case, we observe oscillations of $P_{23}$ between Bosonic and Fermionic 
behaviors due to the competition between entangled electron pair and
the quasiparticles. This can be easily checked experimentally by changing
the gate voltage. 

In summary, we have proposed an entangled electron pair HBT experiment by
using the three-terminal N-S-S hybrid mesoscopic system. When the external
voltage is less than the gap energy, only two-electron current is present.
The cross correlation is found to be positive, which demonstrates that this 
entangled electron pair looks like the composite Boson and tends to bunch 
in cluster. However, when the external voltage is larger than the gap 
energy the quasiparticle will participate the transport which gives
the Fermionic contribution to the cross correlation. As the result of
competition between Andreev reflection process and the other tunneling
process involving quasiparticles, the cross correlation can be either
positive or negative depending on which one dominates. For the strong
coupling case and at large external voltage, the cross correlation 
function changes sign as one varys the gate voltage which controls 
the position of the resonant level.

\section*{Acknowledgments}
We gratefully acknowledge support by a RGC grant from the SAR Government of 
Hong Kong under grant number HKU 7091/01P and a CRCG grant from the
University of Hong Kong.

\bigskip

\noindent{$^{a)}$ Electronic mail: jianwang@hkusub.hku.hk}

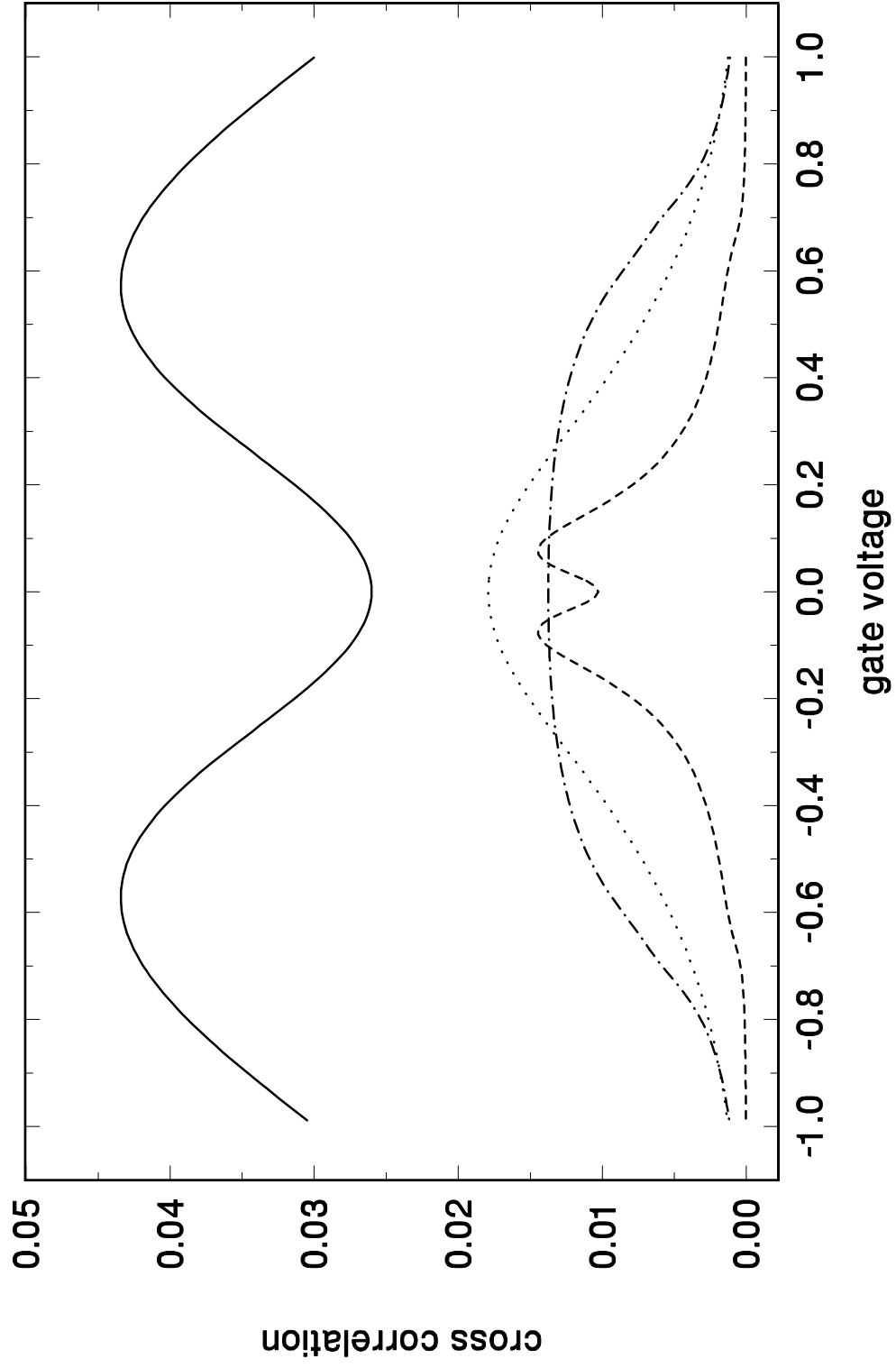
\begin{figure}
\caption{
The cross correlation $P_{23}$ versus gate voltage at $eV=0.6$ for
different coupling parameters: (1). $\Gamma_1=0.8$ and $\Gamma_2=0.8$ 
(solid line); (2). $\Gamma_1=0.8$ and $\Gamma_2=0.1$ (dotted line);
(3). $\Gamma_1=0.1$ and $\Gamma_2=0.8$ (dot-dashed line); (4). 
$\Gamma_1=0.1$ and $\Gamma_2=0.1$ (dashed line). 
}
\end{figure}

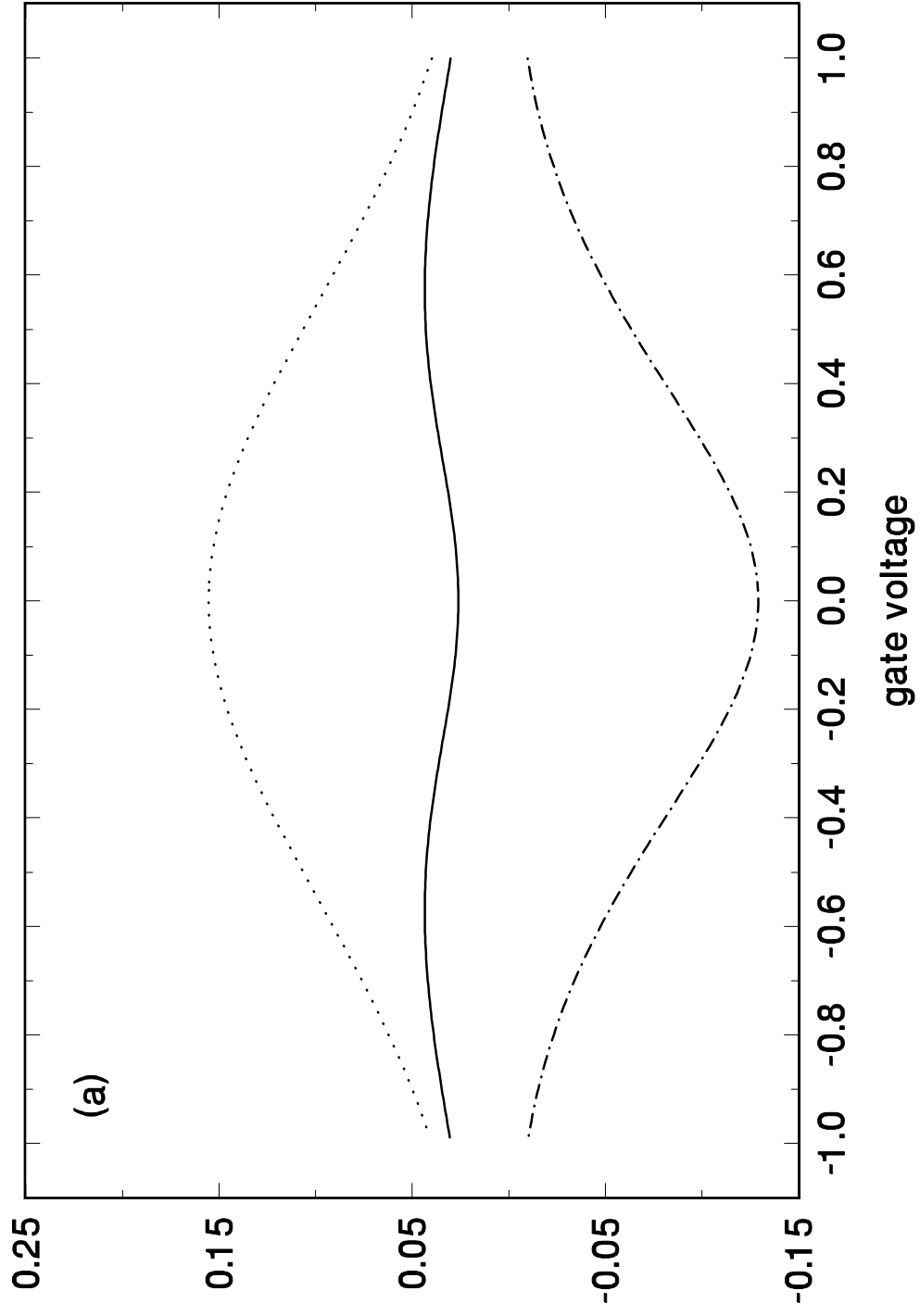
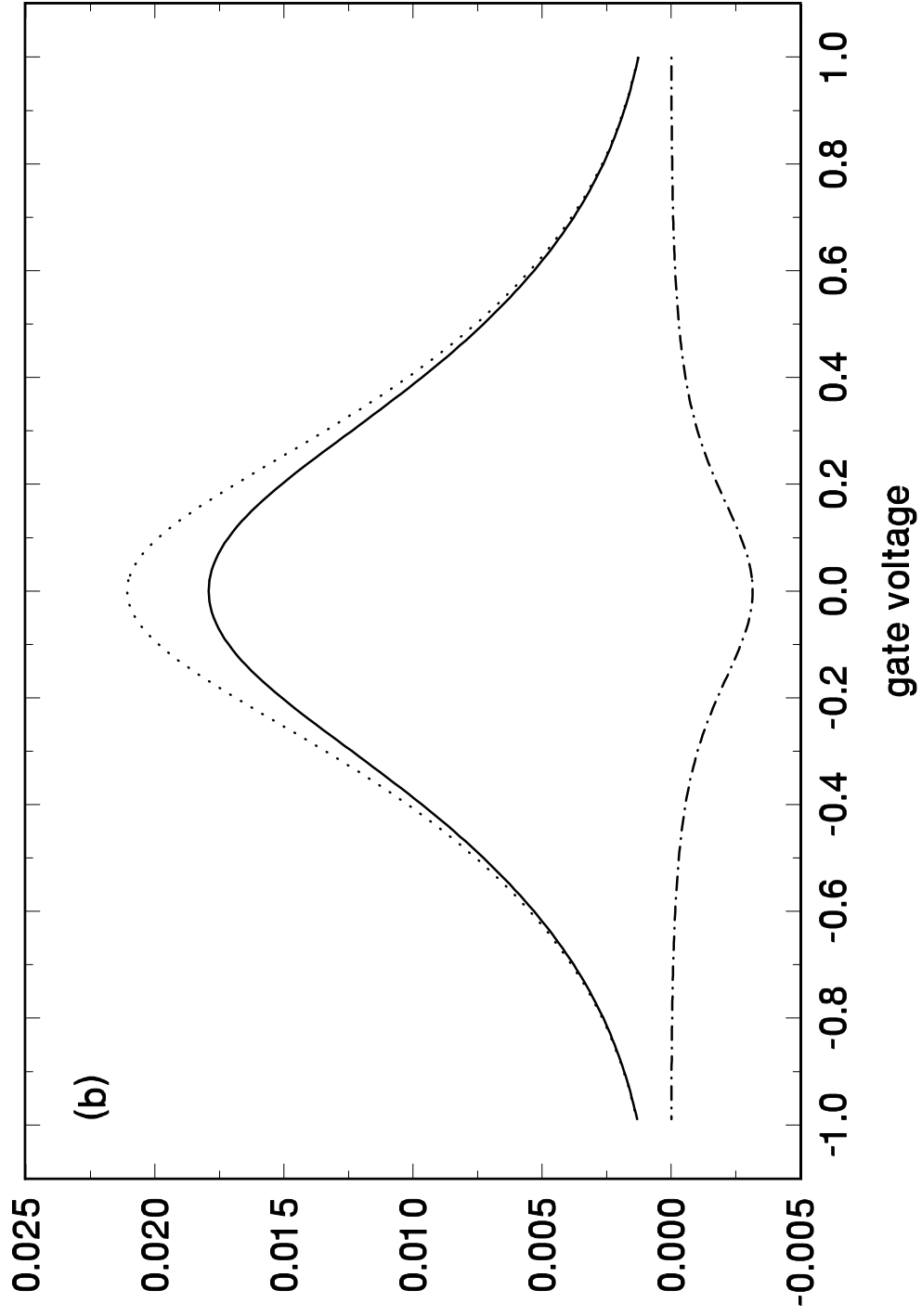
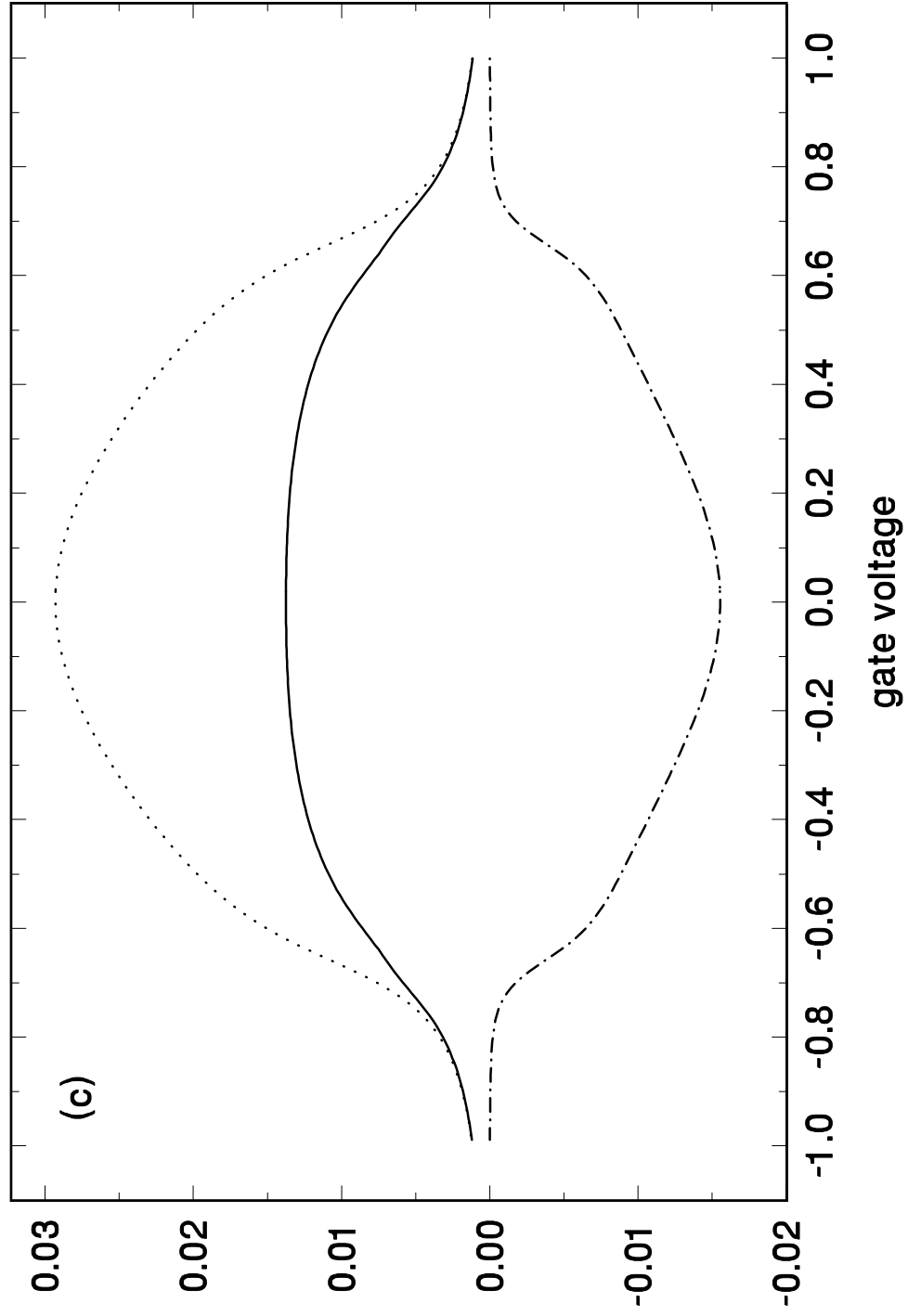
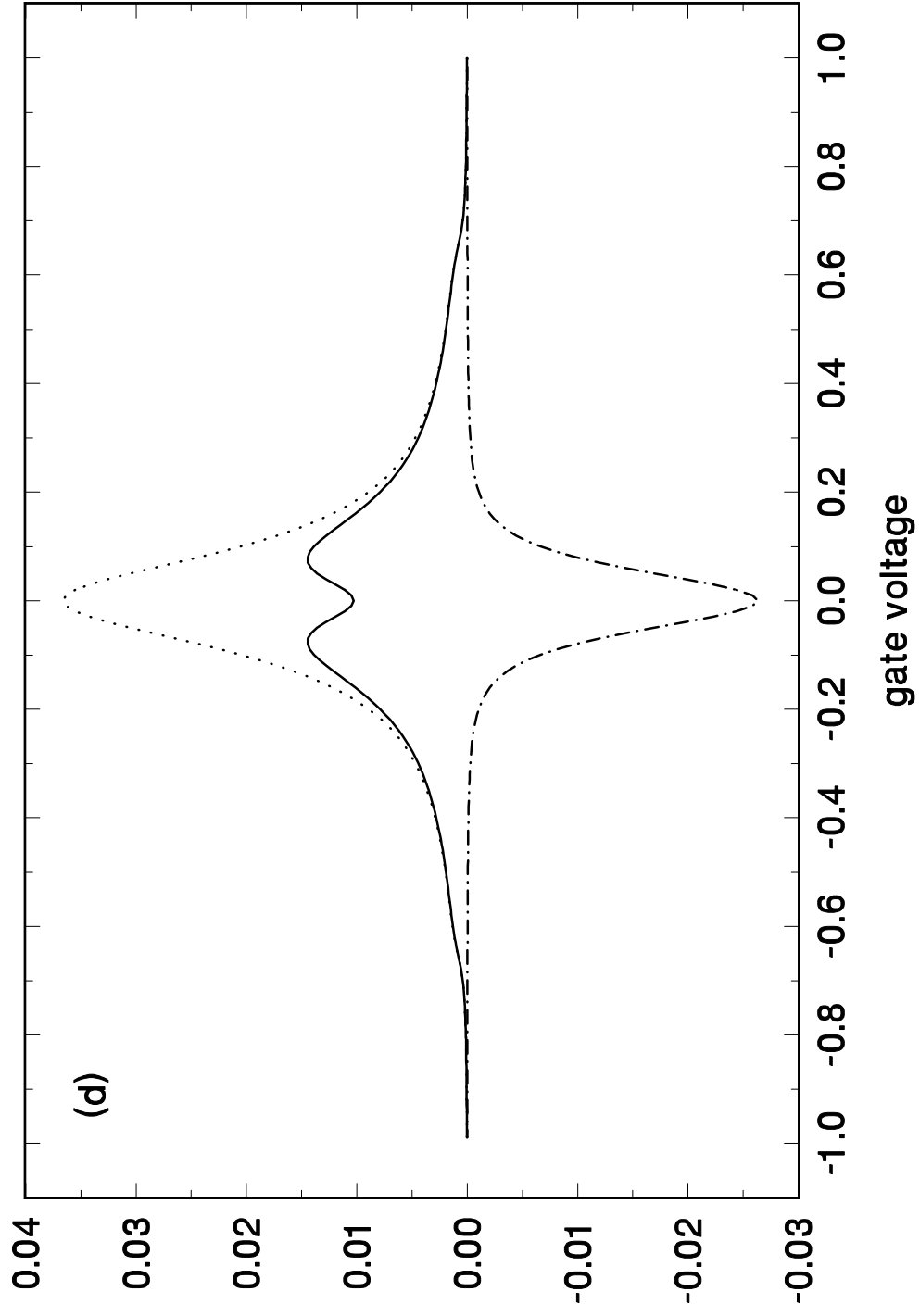
\begin{figure}
\caption{
The contribution of $F_1$ (dotted line) and $F_2$ (dot-dashed line) 
to the cross correlation $P_{23}$ (solid line) versus gate voltage at 
$eV=0.6$. (a). $\Gamma_1=0.8$ and $\Gamma_2=0.8$; (b). $\Gamma_1=0.8$ 
and $\Gamma_2=0.1$; (c). $\Gamma_1=0.1$ and $\Gamma_2=0.8$; 
(d). $\Gamma_1=0.1$ and $\Gamma_2=0.1$. 
} 
\end{figure} 

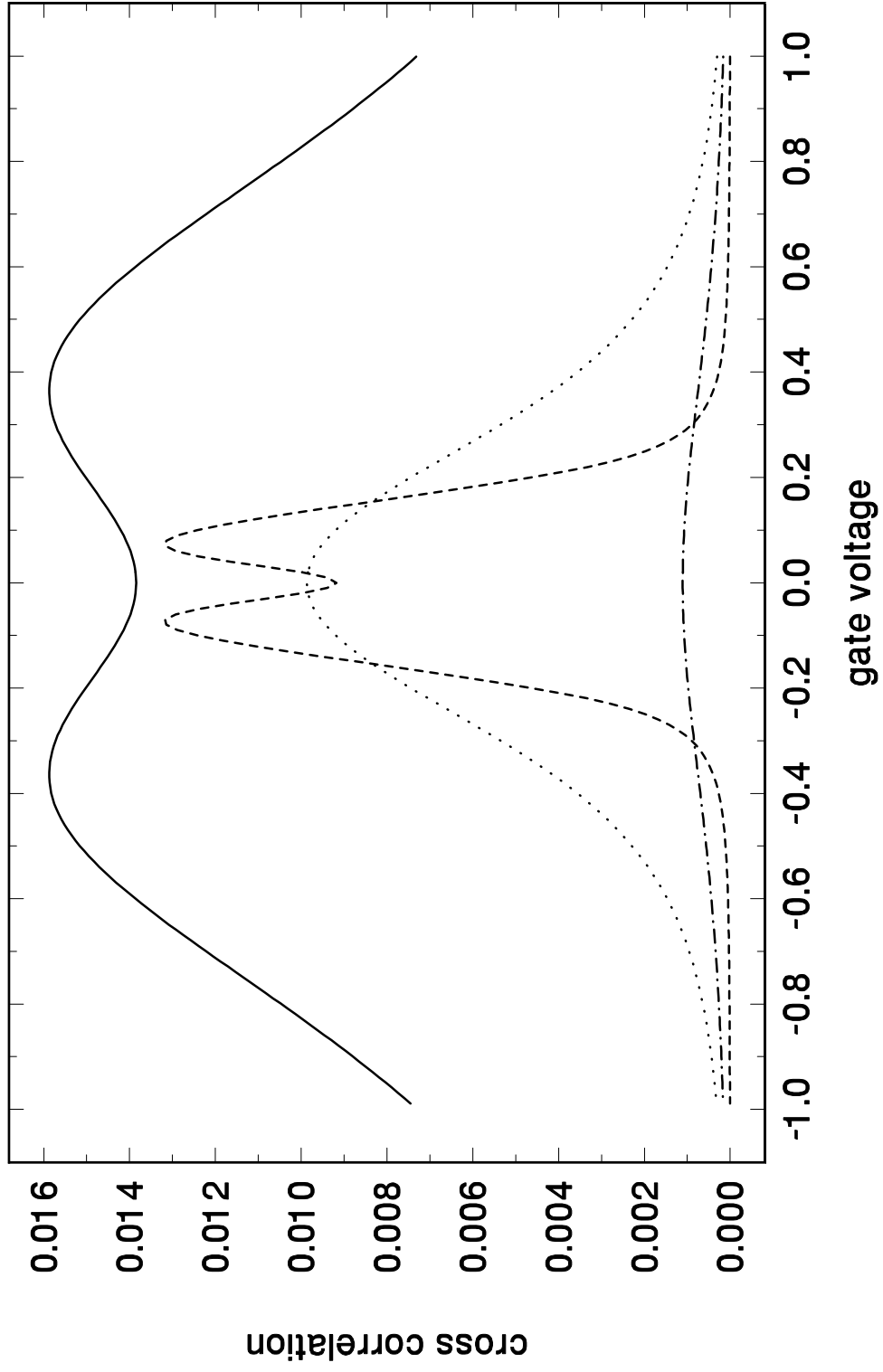
\begin{figure} 
\caption{ 
The cross correlation $P_{23}$ versus gate voltage at $eV=0.2$.
The coupling parameters and corresponding symbols are the same as Fig.1.
}
\end{figure}

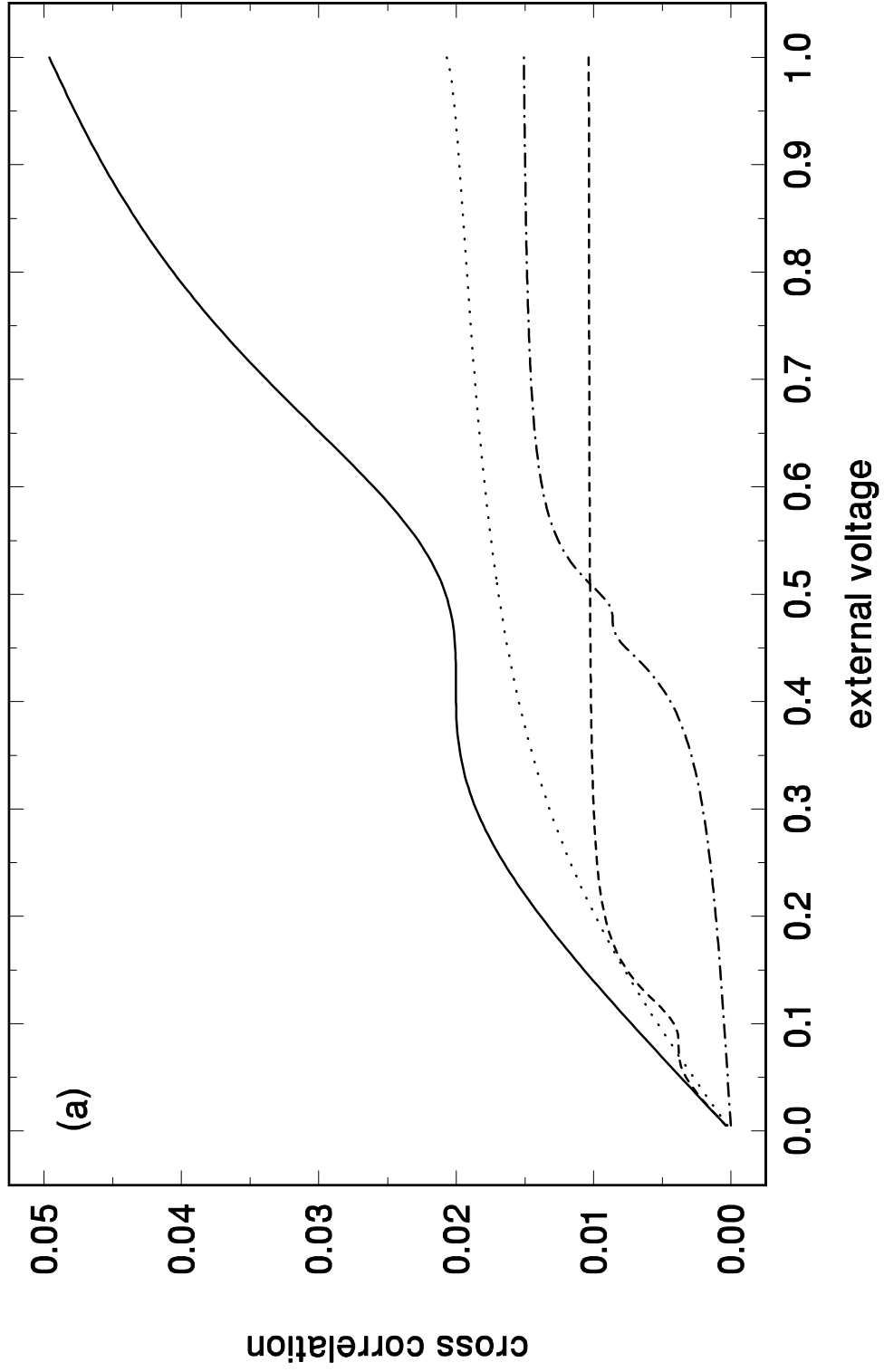
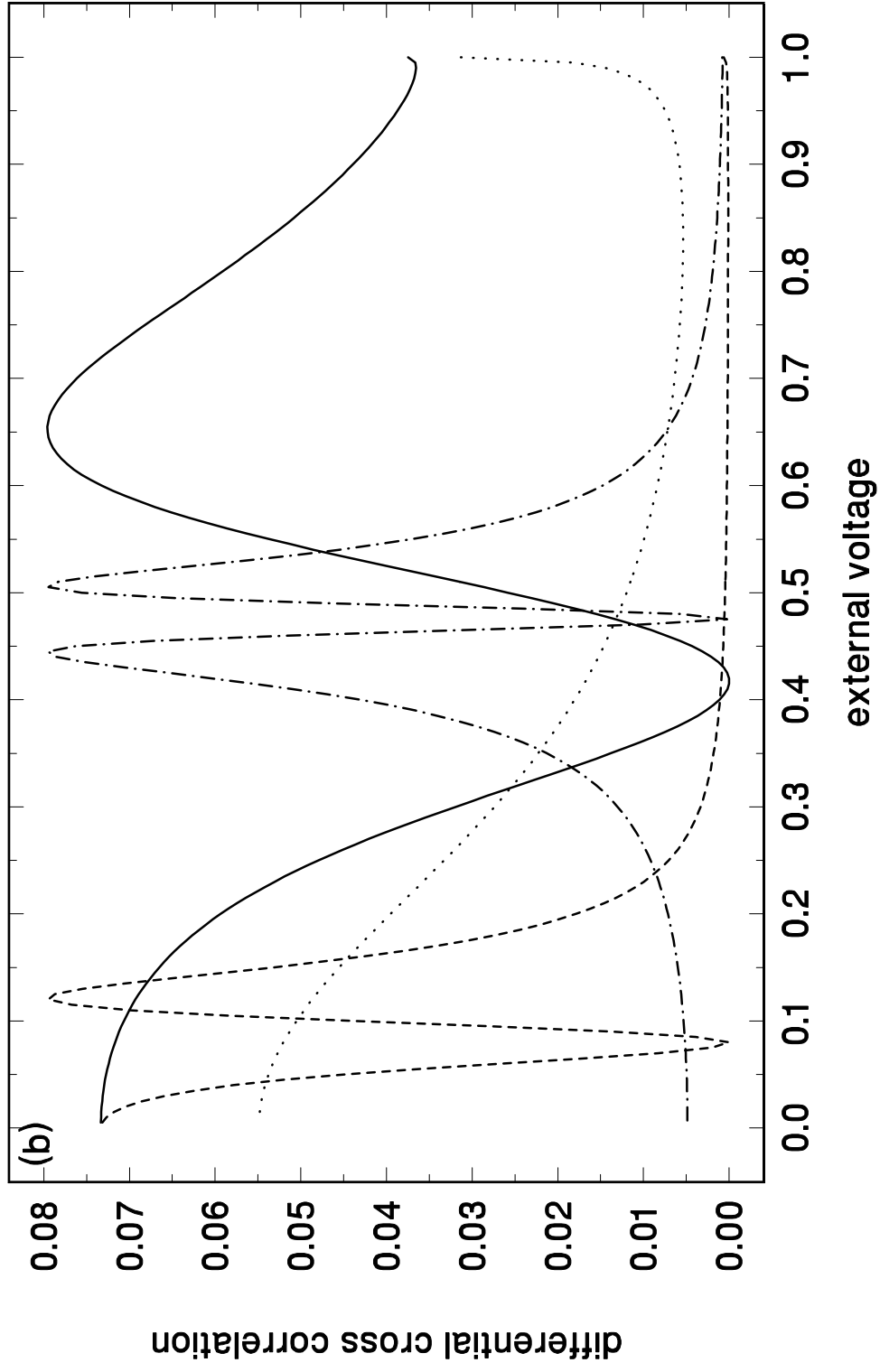
\begin{figure} 
\caption{ 
(a). The cross correlation versus external voltage at $\epsilon_0=0.0$.
(b). The differential cross correlation versus external voltage at 
$\epsilon_0=0.0$. 
The coupling parameters and corresponding symbols are the same as Fig.1.
}
\end{figure}

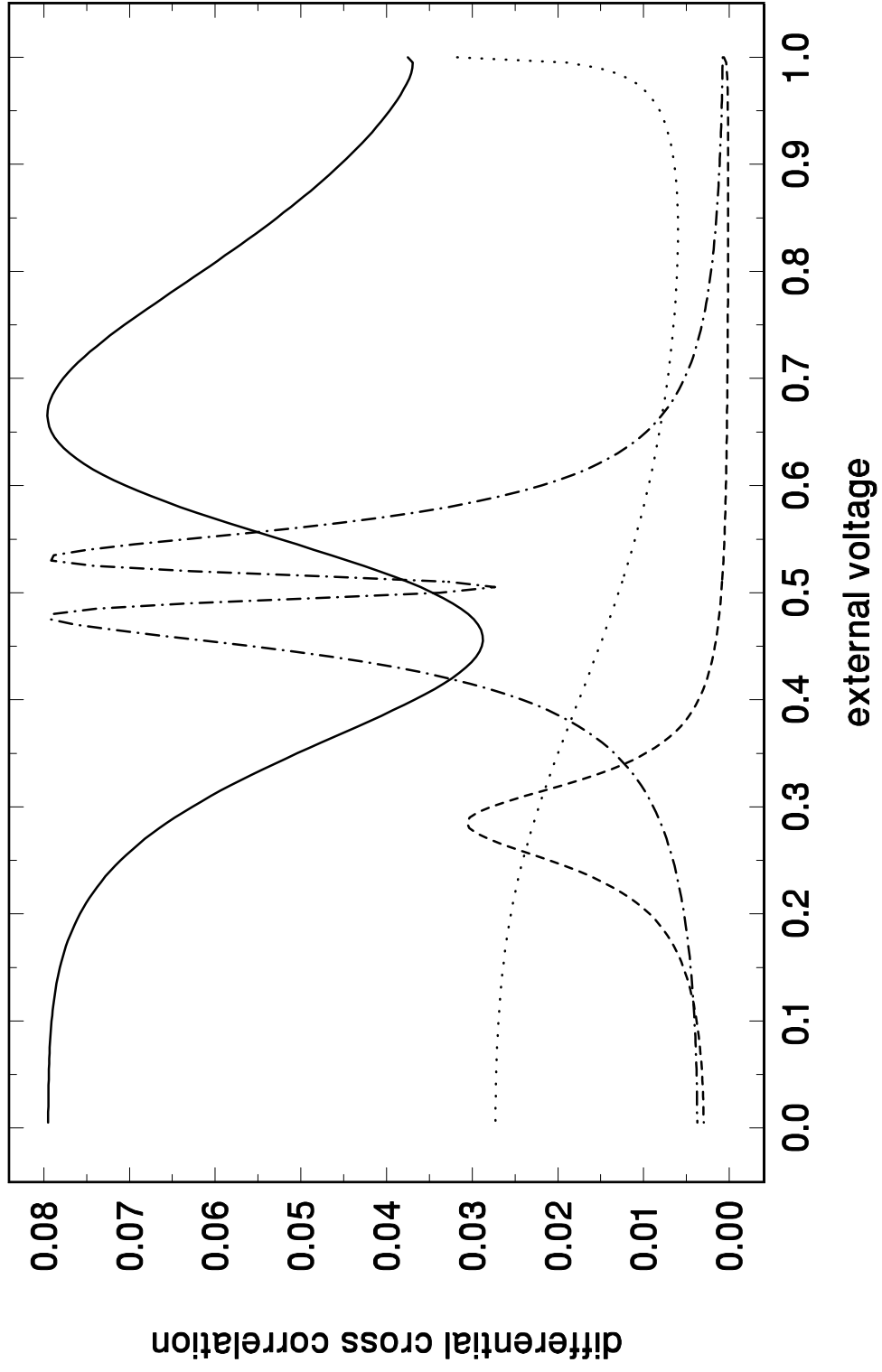
\begin{figure} 
\caption{ 
The differential cross correlation versus external voltage at 
$\epsilon_0=0.3$. 
The coupling parameters and corresponding symbols are the same as Fig.1.
}
\end{figure}

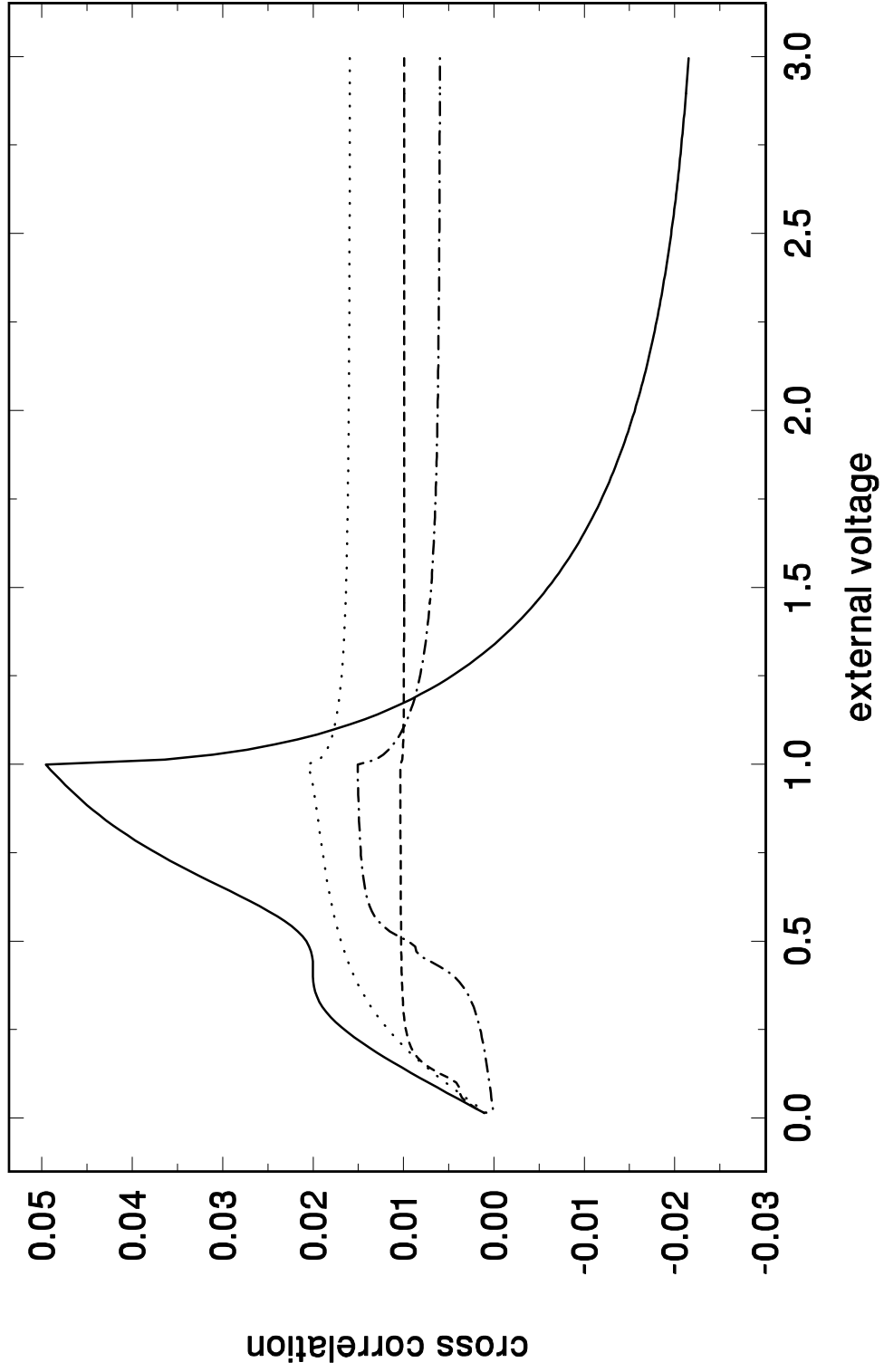
\begin{figure} 
\caption{ 
The cross correlation versus external voltage at $\epsilon_0=0.0$. 
The coupling parameters and corresponding symbols are the same as Fig.1.
}
\end{figure}

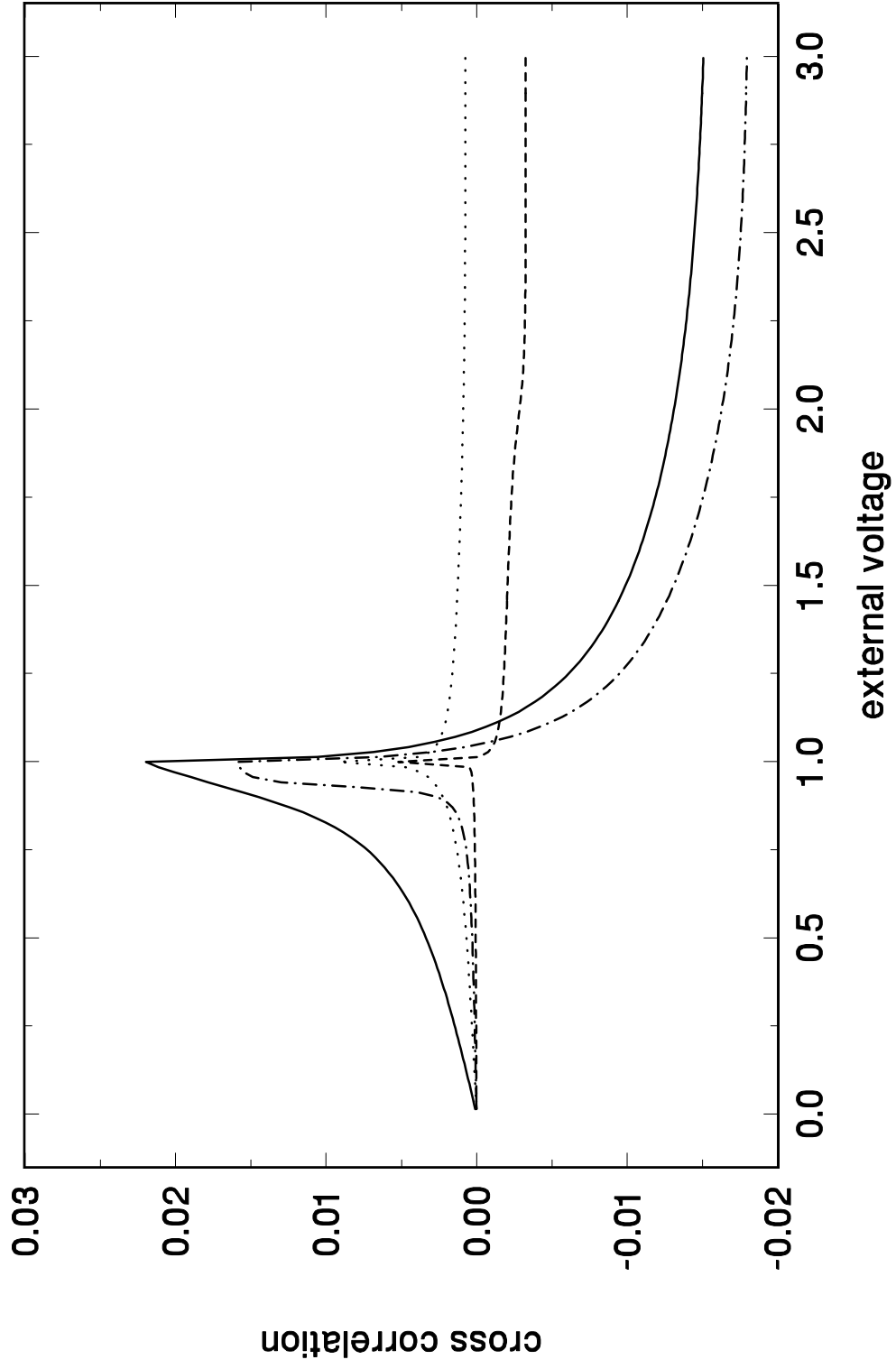
\begin{figure} 
\caption{ 
The cross correlation versus external voltage at $\epsilon_0=2.0$. 
The coupling parameters and corresponding symbols are the same as Fig.1.
For illustration purpose, we have multiplied the cross correlation
by a factor of 10 for the dotted line, 5 for the dot-dashed line, and 
50 for the dashed line.
}
\end{figure}

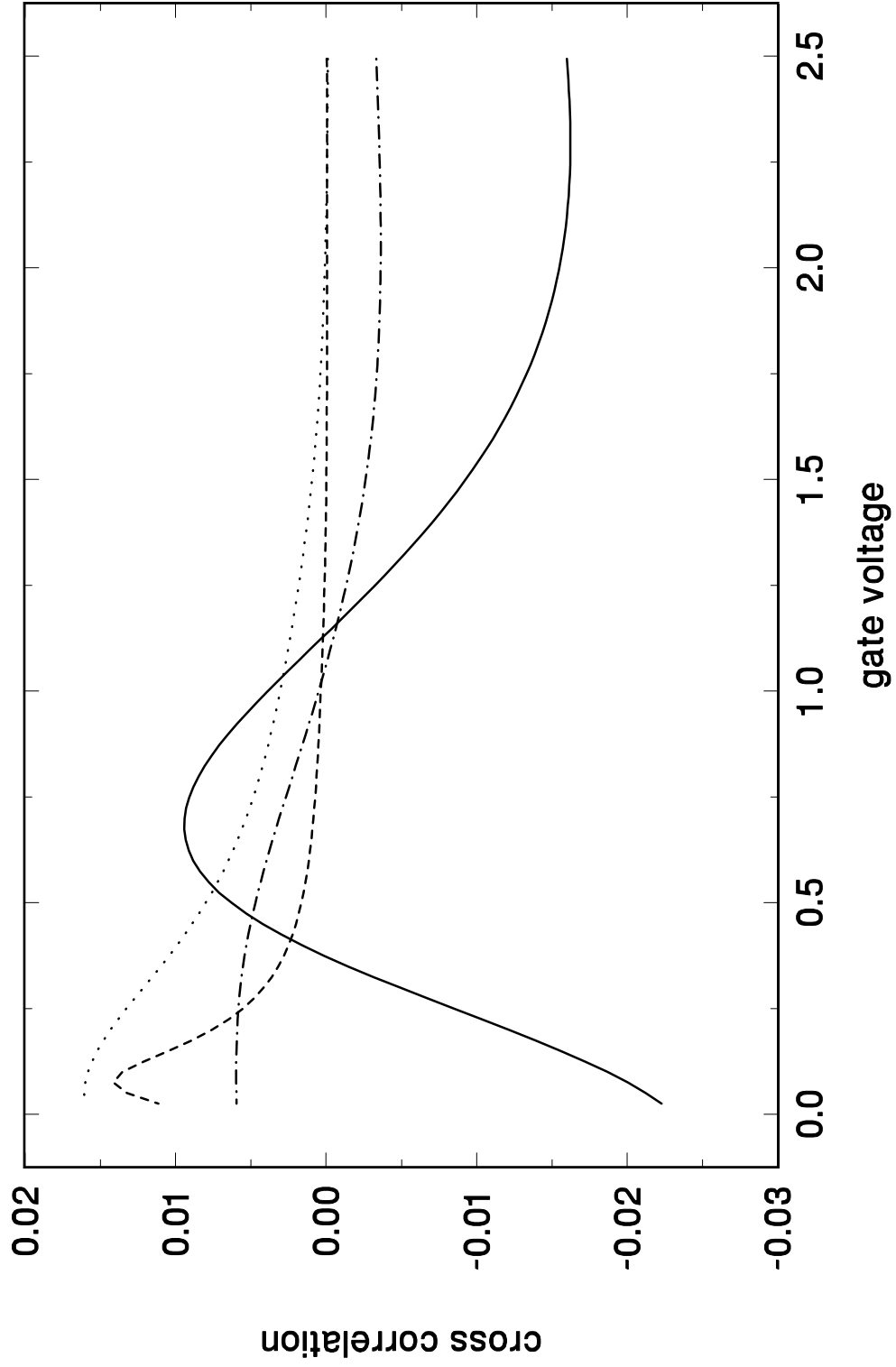
\begin{figure} 
\caption{ 
The cross correlation versus gate voltage at $eV=4.0$. 
The coupling parameters and corresponding symbols are the same as Fig.1.
}
\end{figure}

\begin{thebibliography}{00}
\bibitem{foot1}
Here the density of state in the Green's function has been absorbed in
the linewidth function $\Gamma$.
\end{thebibliography}
\end{document}